\documentclass[aps,prfluids,showpacs,superscriptaddress,preprint,floatfix,amsmath,longbibliography]{revtex4-1}
\usepackage{amsfonts}
\usepackage{amssymb}
\usepackage{gensymb}
\usepackage{xcolor}
\usepackage{graphicx}
\usepackage{wasysym}
\usepackage[textsize=footnotesize]{todonotes}
\usepackage{siunitx}
\usepackage{float}
\usepackage{lmodern}
\usepackage{bm}
\usepackage{array}

\definecolor{darkblue}{rgb}{0,0,0.6}
\definecolor{darkred}{rgb}{0.6,0,0}
\usepackage[colorlinks=true,urlcolor=darkblue, citecolor=darkblue, linkcolor=darkred, hyperfootnotes=false]{hyperref}

\newcommand{\iP}{\tilde{\Pi}}
\newcommand{\iS}{\tilde{\sigma}}

\begin{document}
%\graphicspath{{./Figs/}}

\title{Transition from granular to Brownian suspension : an inclined plane experiment}

\author{Alice Billon}
\affiliation{Aix Marseille Univ, CNRS, IUSTI, Marseille 13013, France}
\affiliation{Gulliver, UMR CNRS 7083, ESPCI Paris, PSL Research University, 10 rue Vauquelin, 75005 Paris, France}
\author{Yo\"el Forterre}
\affiliation{Aix Marseille Univ, CNRS, IUSTI, Marseille 13013, France}
\author{Olivier Pouliquen}
\affiliation{Aix Marseille Univ, CNRS, IUSTI, Marseille 13013, France}
\author{Olivier Dauchot}
\email{olivier.dauchot@espci.fr}
\affiliation{Gulliver, UMR CNRS 7083, ESPCI Paris, PSL Research University, 10 rue Vauquelin, 75005 Paris, France}
\begin{abstract}
We experimentally revisite the flow down an inclined plane of dense granular suspensions, with particles of sizes in the micron range, for which thermal fluctuations cannot be ignored. Using confocal microscopy on a miniaturized set-up, we observe that, in contrast with standard granular rheology, the flow profiles strongly depend on the particles size. Also, suspensions composed of small enough particles flow at infinitesimal inclinations. From the velocity measurements, an effective rheology is extracted in terms of a friction coefficient as a fonction of the dimensionless shear rate (the viscous number), and of the particle pressure normalized by the thermal pressure. Inspired by a previous work \cite{ikeda_unified_2012}, a phenomenological model based on the sum of a thermal contribution describing the glass transition and an athermal contribution capturing the jamming transition is developed, which reproduces well the experimental observations. The model predicts the existence of a glassy friction angle lower than the granular athermal friction angle, a signature of the glass transition in the framework of a pressure imposed rheology. 
\end{abstract}

\maketitle
%When flowing at constant packing fraction $\phi$ at low particulate Reynolds number, they obey a quasi-Newtonian rheology. if particles interact mainly through hydrodynamic forces and simple Coulomb friction
%in the viscous regime through hydrodynamic and Coulomb contact forces mainly
%and as long as inertia is neglected and forces other than hydrodynamic interactions or simple Coulomb friction are not considered, 

Suspensions composed of large enough particles, so that thermal fluctuations can be ignored, are called granular suspensions. When flowing at constant packing fraction $\phi$, and as long as inertia is neglected and the particles interact only via hydrodynamic forces and simple Coulombian solids contacts, they obey a Newtonian rheology. The shear stress $\sigma$ and the normal granular stress $\Pi$ are proportional to the shear rate $\dot \gamma$ (the only timescale in the problem),  and the shear and normal viscosities $\eta_s(\phi)$ and $\eta_n(\phi)$ are function of the volume fraction, which diverge at the jamming transition, when $\phi$ approaches the maximum volume fraction $\phi_J$ \cite{denn2014rheology,bonnoit_inclined_2010,boyer_unifying_2011,guazzelli_rheology_2018}. An alternative description consists in considering the rheology at constant imposed granular pressure $\Pi$, $\phi$ being free to adjust ~\cite{jop_constitutive_2006,forterre2008flows}, a situation typically encountered in avalanche flows under gravity. Within this framework, the constitutive laws are given by the friction coefficient $\mu = \sigma/\Pi$ and the volume fraction $\phi$ as unique functions of the so-called viscous number $J = \eta_s \dot\gamma/\Pi$~\cite{cassar2005submarine}, where $\eta_s$ is the viscosity of the suspending fluid~\cite{boyer_unifying_2011,guazzelli_rheology_2018}. The jamming transition occurs in the limit of vanishing $J$, where  the packing fraction reaches $\phi_J$ and $\mu$ converges to a finite value $\mu_J$. The latter denotes the existence of an angle of repose $\theta_J = \tan^{-1}(\mu_J)$, below which the suspension does not flow under gravity. For frictionless spherical particles, as those considered in this paper, the jamming packing fraction is $\phi_J\simeq 0.64$ while the jamming friction coefficient  is $\mu_J\simeq 0.1$, corresponding to a pile angle $\theta_J\simeq 5-6^{\circ}$~\cite{peyneau2008frictionless}. 

 Suspensions composed of small particles, the dynamics of which is sensitive to thermal fluctuations, are called Brownian suspensions. Thermal fluctuations introduce an additional timescale, whose comparison with the shear rate defines the P\'eclet number and therefore modifies the dimensional analysis underlying the above description \cite{trulsson_athermal_2015,wang_constant_2015}. Dense Brownian suspensions exhibit a glass transition at a packing fractions $\phi \sim \phi_G$~\cite{pusey_phase_1986}, which is marked by a divergence, or if not, a very sharp increase of the structural relaxation time of the suspension. This induces a divergence of the viscosity of the suspension and the emergence of a thermal yield stress  when $\phi$ is greater than $\phi_G$~\cite{cheng2002nature,siebenburger2009viscoelasticity}. 

Despite the similarity of the flowing properties of granular and Brownian suspensions, it has been  shown that both the glass and jamming physics impact the flow curves over distinct stress scales and time scales~\cite{ikeda_unified_2012}. Eventually, a simple additive model, where the shear stress is the sum of the glass and jamming contributions, was shown to capture both numerical and experimental data of a variety of dense suspensions flowing at imposed volume fraction in the thermal crossover~\cite{ikeda_unified_2012,ikeda_disentangling_2013}. 

However, we are still lacking a complete description of the flow of Brownian suspensions when the granular pressure is imposed rather than the packing fraction, a situation of significant interest for flow and transport of colloids or agitated particles under gravity~\cite{peshkov2016active,berut2018gravisensors,berut_brownian_2019,houssais2021athermal}. The presence of thermal agitation introduces an additional dimensionless number  $\iP=\Pi d^3/k_BT$ comparing the confining pressure to a thermal pressure, with $d$ the particle diameter, $k_B$ the Boltzmann constant and $T$ the absolute temperature. The rheology is then expected to be described by a friction coefficient $\mu=\mu(J,\iP)$ and a volume fraction $\phi=\phi(J,\iP)$ function of both $J$ and $\iP$.  Numerically, Trulsson et al  \cite{trulsson_athermal_2015} used discrete element simulations to study the pressured imposed rheology of thermal suspensions and have shown that it is analogous to soft athermal particles,  in which a soft repulsive interaction  mimic the random thermal force, except at low P\'eclet number. Using Brownian particle simulation Wang and Brady \cite{wang_constant_2015} have studied in details the pressure imposed rheology and measured the $\mu(J,\iP)$  and $\phi(J,\iP)$ laws. They have shown that the critical friction coefficient in the quasi-static regime is affected by the thermal agitation and drops to zero for strong agitation.  Experimentally, avalanche flows of micrometer sized particles in rotating drums have been studied \cite{berut_brownian_2019}, showing that for small enough grains the avalanche does not stop at a finite repose angle, as it does for granular suspensions, but slowly creeps until the pile free surface becomes horizontal. This transition towards a vanishingly small pile angle remains largely unexplored and the possible link between the observed creep and the glassy dynamics remains elusive.

In the present work we bring experimental and theoretical evidences for the transition between thermal and athermal suspensions. To do so, we experimentally study the flow of a layer of micrometer sized particles down an inclined plane, a classical configuration investigated in the granular regime~\cite{bonnoit_inclined_2010, pailha_initiation_2008, perrin_thin_2021}. The main advantage of this configuration is that steady uniform flows are easily achieved, where the friction coefficient and the granular pressure distribution are known, enabling the extraction of the rheology. We use confocal microscopy to access the velocity profiles in a steady flow regime. By changing the size of the particles, we investigate the role of thermal agitation on the velocity profile and rheology, showing that the friction coefficient vanishes for sufficiently small grains. To describe the results,  a generalized granular rheology in the presence of thermal fluctuations is derived in terms of the two laws $\mu(J,\iP)$  and $\phi(J,\iP)$,  by extending to the normal stress the additive model proposed for the shear stress by Ikeda et al \cite{ikeda_disentangling_2013}.  The prediction of the model is compared to the experiments and previous numerical simulations~\cite{wang_constant_2015}.\\

 \begin{figure}[t!]
\begin{center}
\includegraphics[width=12 cm]{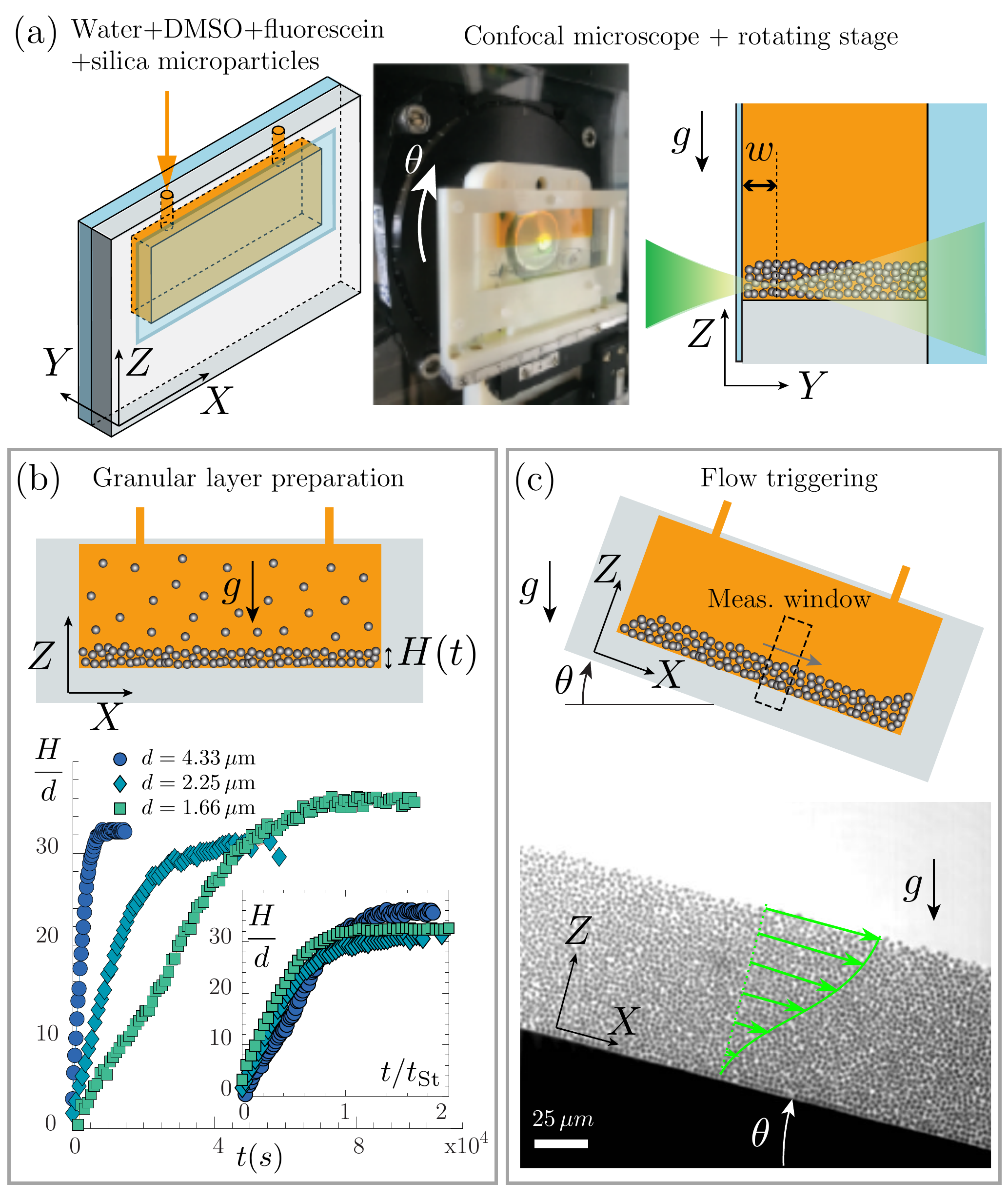}\\
\end{center}
\vspace{-2mm}
\caption{{\bf Experimental protocol.} a) experimental setup:  the bi-disperse suspension (see Table \ref{table1}) is introduced in a cell, made of a PDMS sheet (grey) sealed between a glass slide and a cover slip (blue). The cell is fixed to a translation stage and a rotation stage on an inclined confocal microscope, taking images a distance $w=10d$ from the cover slip. (b) preparation: after mixing, the particles sediment and form a uniform layer of height $H(t)$; inset shows $H/D$ as a function of the time rescaled by the Stokes time (see text). (c) inclination: at $t=0$ the cell is inclined at an angle $\theta$ and velocity profile are obtained  using PIV (picture obtained for $d=2.25$ $\mu$m at $\theta=15^{\circ}$).}
\vspace{-3mm}
\label{fig:setup}
\end{figure}

\textbf{Materials and Methods} -- The experimental setup and protocol are sketched in figure \ref{fig:setup}. To study the role of thermal fluctuations, different silica beads (Microparticles GmbH, density $\rho_p=1,850$ kg$\cdot$m$^{-3}$) are used  with a mean diameter $d$ varying between 0.91 and 4.3 $\mu$m. To prevent crystallization, for each mean size $d$, we use a suspension made of an equal mixture of two batches of particles with slightly different sizes as indicated in the table~\ref{table1}. To enable confocal imaging, the particles are immersed in an index-matched liquid composed of water, dimethylsylfoxide and fluorescein (water mass content 14--40\%wt depending on the sample), giving a suspending fluid's  density $\rho_s=1,100$ kg$\cdot$m$^{-3}$, viscosity $\eta_s=3.1-3.7$ mPa$\cdot$s and refractive index $n=1.42-1.46$. It is important to emphasize that the silica beads in water behave like frictionless particles \cite{clavaud2017revealing,berut_brownian_2019} due to the presence of a short-range repulsion force of electrostatic origin between the negatively charged surfaces of the particles, which is large enough to sustain the typical weight of the layer of particles investigated in this study. 

The cell is a long rectangular cavity moulded in PDMS, with a length $52$ mm ($X$-direction), a height $14$ mm ($Z$-direction)  and a thickness $1.5$ mm ($Y$-direction), sealed between a glass slide and a glass coverslip (Fig.~\ref{fig:setup}a). The cell is fixed on a translation stage and a rotation stage  attached to an inclined confocal microscope, whose optical axis is perpendicular to gravity.
\begin{table}[h!]
\begin{center}
\begin{tabular}{ |c|c|c|c|c|c|c| } 
 \hline
 sample 	& $d_1 (\mu{\rm m})$ & $d_2 (\mu{\rm m})$ & %$d=\frac{d_1+d_2}{2} 
 $d={\left(\frac{d_1^4+d_2^4}{2}\right)}^{1/4} (\mu{\rm m})$ 	&   $\iP_0$  & symbol \\ 
 1    	     	& 0.83                         & 0.98                         &      0.91                                        &      1.26        	   &   $\Circle$	  \\ 
 2    		& 1.53    			  & 1.76   			   & 	   1.66 					&   13.9		   & 	$\Diamond$ \\
 3    		& 2.12    			  & 2.36   			   & 	   2.25  					&   47.7		   &	$\square$	  \\
 4    		& 3.97     			  & 4.62   			   &      4.33  					& 649		   &	$\triangledown$    \\
 \hline
\end{tabular}
\end{center}
\vspace{-3mm}
\caption{Particle sizes, dimensionless weight $\iP_0= \delta \rho g d^4/(k_BT)$, with $\delta\rho=\rho_p-\rho_s=750$~kg$\cdot$m$^{-3}$, $g=9.81$ m$\cdot$s$^{-2}$ the intensity of gravity, $k_BT=4 \times 10^{-21}$ J$\cdot$K$^{-1}$ and symbol for the four suspension samples used. The mean diameter $d$ for the bidisperse mixture is defined such that $\iP_0(d)=\frac{1}{2}(\iP_0(d_1)+\iP_0(d_2))$ .  }
\label{table1}
\end{table}
 Once the cell is filled with the suspension, it is agitated to mix the suspension, before letting the particles sediment in a horizontal position (Fig.~\ref{fig:setup}b). The amount of particles in the cell is chosen such that the final thickness of the deposit is approximately $H\simeq 30d$. The time evolution of the height of the deposit is recorded during the sedimentation (Fig.~\ref{fig:setup}b), showing that the sedimentation is finished after about twice the Stokes' time $t_{\rm St}= \delta\rho g H_{\rm cell} d^2/18\eta_s$~\cite{lebel_density_1962}, where $H_{\rm cell}=14$ mm is the height of the cell. Once the particles sedimented, we checked the horizontal homogeneity of the deposit along the cell before each flow measurement (the local maximum slope of the sediment is less than 1$^o$ ). The cell is then inclined at an angle $\theta$ (Fig.~\ref{fig:setup}c) and the flow of the suspension is observed by confocal imaging at a distance $w=10d$ from the side wall of the cell. The flow is recorded at regular intervals and the velocity profiles $U(z,t)$ and the height of the deposit $H(t)$ are obtained from a standard PIV analysis. 
 
% , with $\delta\rho = 750$ kg$\cdot$m$^{-3}$ and $\eta_s$ is the viscosity of the index-matching solution 

\begin{figure}[t!]
\begin{center}
\includegraphics[width=12cm]{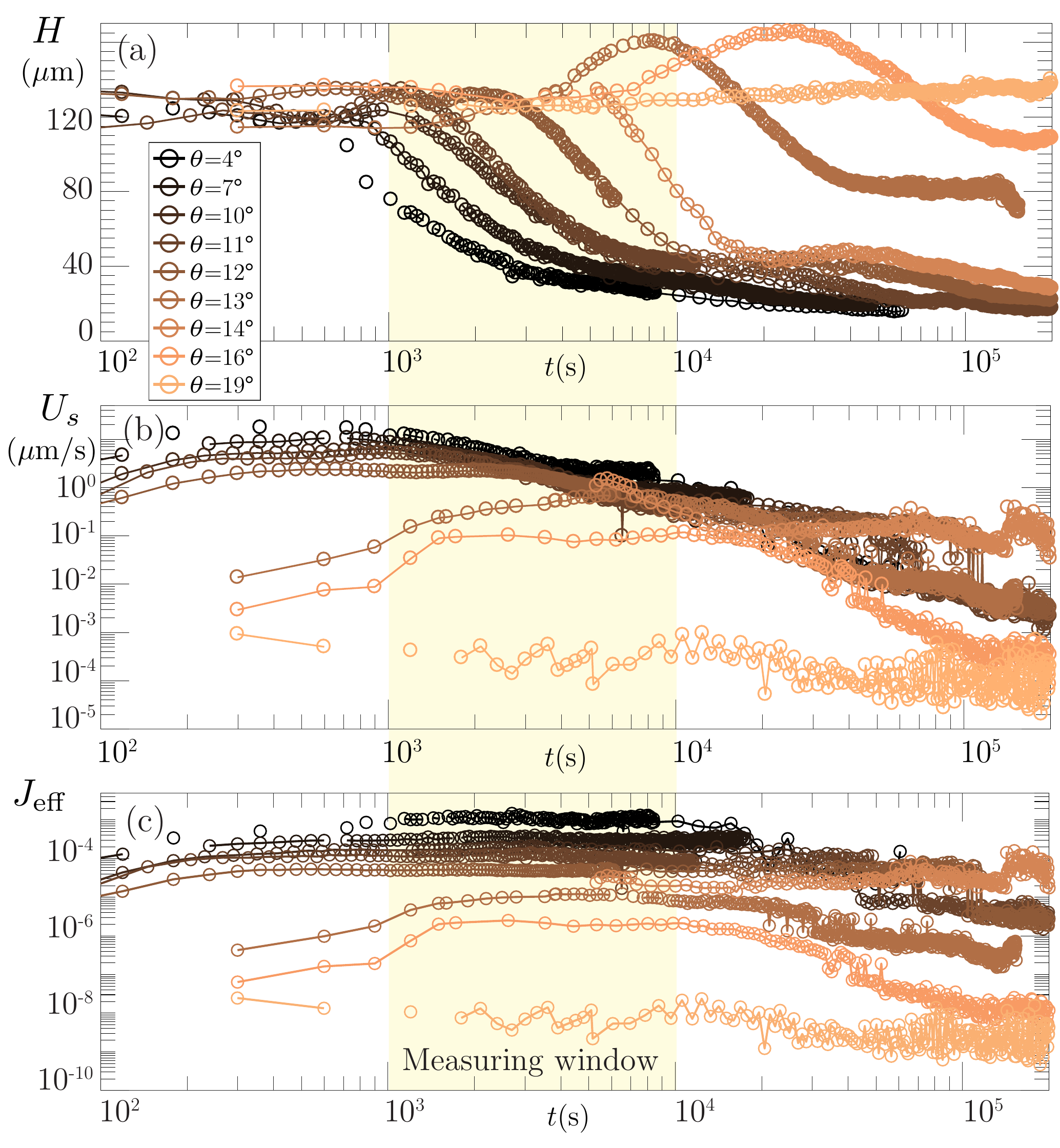}
\end{center}
\vspace{-2mm}
\caption{{\bf Dynamics of the flow:} (a) evolution of the layer thickness $H(t)$, (b) the surface velocity $U_s(t)$, and (c) the effective viscous number $J_{\rm eff}(t)$ for the suspension sample $\#4$ at different inclination angles $\theta$ as indicated in the legend. The yellow region corresponds to the time slot  where measurements are taken to infer the rheology.}
\vspace{-2mm}
\label{fig:dynamics} 
\end{figure}

\textbf{Results} -- Once the flow starts, we monitor the height of the flowing layer, $H(t)$, and the surface velocity, $U_s(t)$, as a function of time (see Fig.~\ref{fig:dynamics}a,b). Both the height of the flowing layer and the surface velocity initially  slightly increases, before continuously decreasing. We also compute in  Fig.\ref{fig:dynamics}c the dimensionless depth averaged shear rate characterized by the effective viscous number $J_{\rm eff}(t)=2 \frac{\eta_s U_s}{H \Pi_b}$ (defined as in \cite{perrin_thin_2021}), where $\Pi_b = <\phi> \delta\rho g H \cos\theta$ is the pressure at the bottom of the layer estimated using $<\phi>=0.6$. We observed that a plateau is reached after an accelerating phase, where $J_{\rm eff}$ is constant ($t\in[10^3, 10^4] s$), ensuring quasi-steady flows. In the following all the measurements are taken and averaged in this regime (yellow region). 

\begin{figure}[t!]
\begin{center}
\includegraphics[width=0.8\columnwidth]{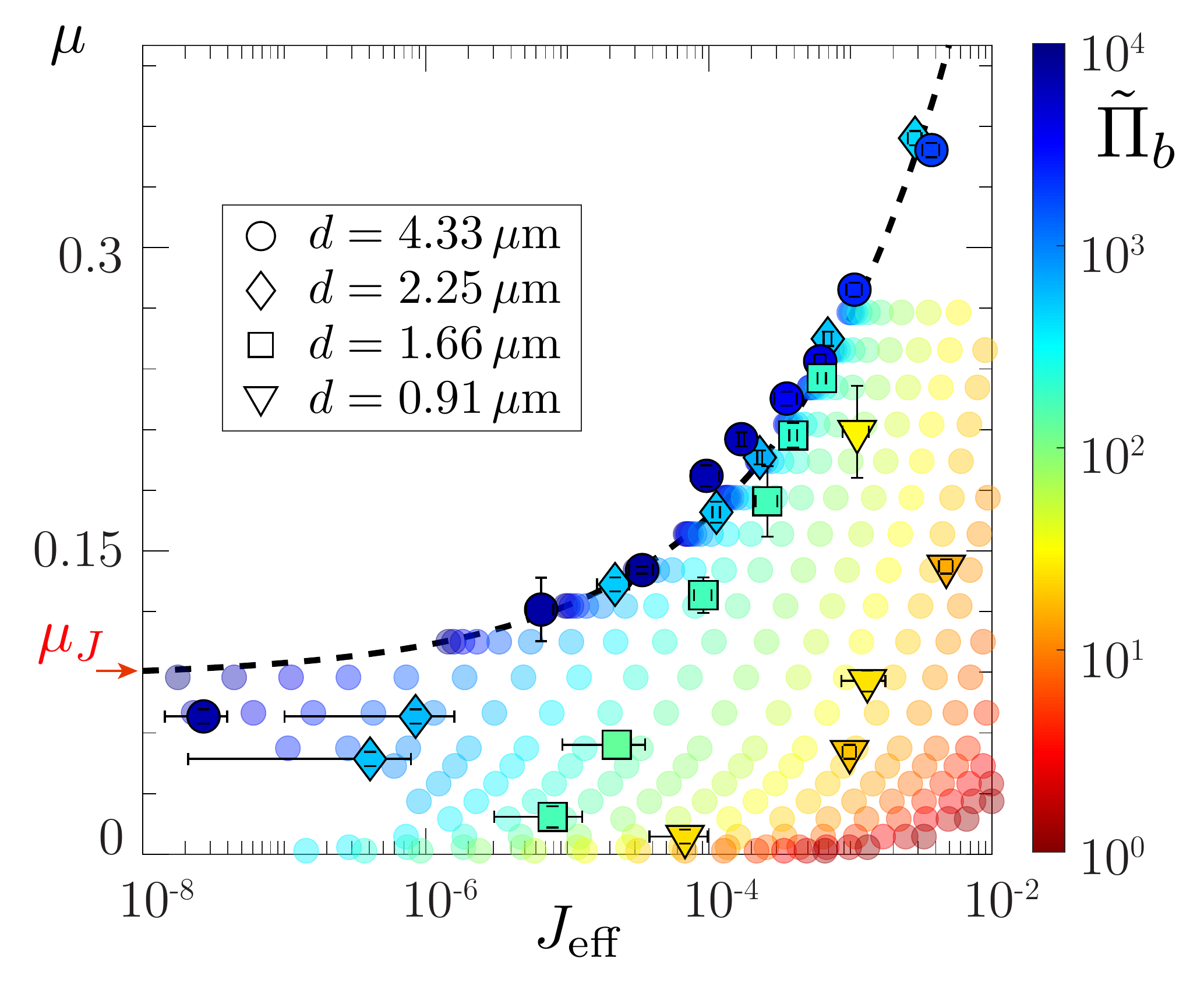}
\vspace{-1mm}
\caption{{\bf Effective rheology:} The $\mu(J_{\rm eff})$ rheology curves with symbols as indicated in Table \ref{table2} and color coding for $\iP_b$. The black dashed line indicates the rheology of an athermal suspension obtained experimentally by Perrin et al~\cite{perrin_thin_2021}. The light colored disks are the prediction of the additive model (see text for details); same color code as for the experimental data.} 
\label{fig:effrheo}
\vspace{-8mm}
\end{center}
\end{figure}

In the inclined plane configuration,  the ratio of the shear stress to the normal stress, i.e the friction coefficient, is constant across the layer for steady uniform flows, being equal to the tangent of the inclination  $\mu = \tan\theta$. We can then infer an effective rheology by plotting $\mu$ as the function of the steady value of the effective viscous number $J_{\rm eff} = \left<J_{\rm eff}(t)\right>_{t\in[10^3, 10^4] }$ for all experimental runs. In Fig.~\ref{fig:effrheo}, we have systematically plotted $\mu(J_{eff})$ for different particle sizes and inclination angles varying from 20$^{\circ}$ to 1.5$^{\circ}$, the color encoding for the dimensionless bottom pressure $\iP_b=\Pi_b d^3/k_BT$. Large value of $\iP_b$ (dark blue) corresponds to the limit of athermal suspensions, whereas small values (red) corresponds to Brownian suspensions. In the limit of athermal suspensions, we expect the data to collapse on a single curve $\mu(J_{eff})$, as the viscous number is the only relevant parameter. This is the case for the two suspensions made of the largest particles, which collapse well on the athermal curve obtained for much larger particles ($d=25\,\mu$m) by Perrin et al \cite{perrin_thin_2021}. However, we note that for very low values of $J_{eff}$, the friction coefficient measured with our particles is systematically smaller than the athermal one $\mu_J\approx0.09$. This means that even with our biggest particles $d=4.3 \, \mu$m, slow flows take place for angles below the athermal granular angle of repose. 
 The deviation becomes  more severe when considering the two suspensions composed of the smallest particles (green and yellow symbols). In that case, it is clear that the rheology departs from the athermal one over the entire range of viscous number. The friction coefficient seems to vanish when decreasing the viscous number, indicating that the suspension flows even for infinitesimal angles, as reported for avalanches in microdrums~\cite{berut_brownian_2019}.

\begin{figure}[t!]
\begin{center}
\includegraphics[width=12cm]{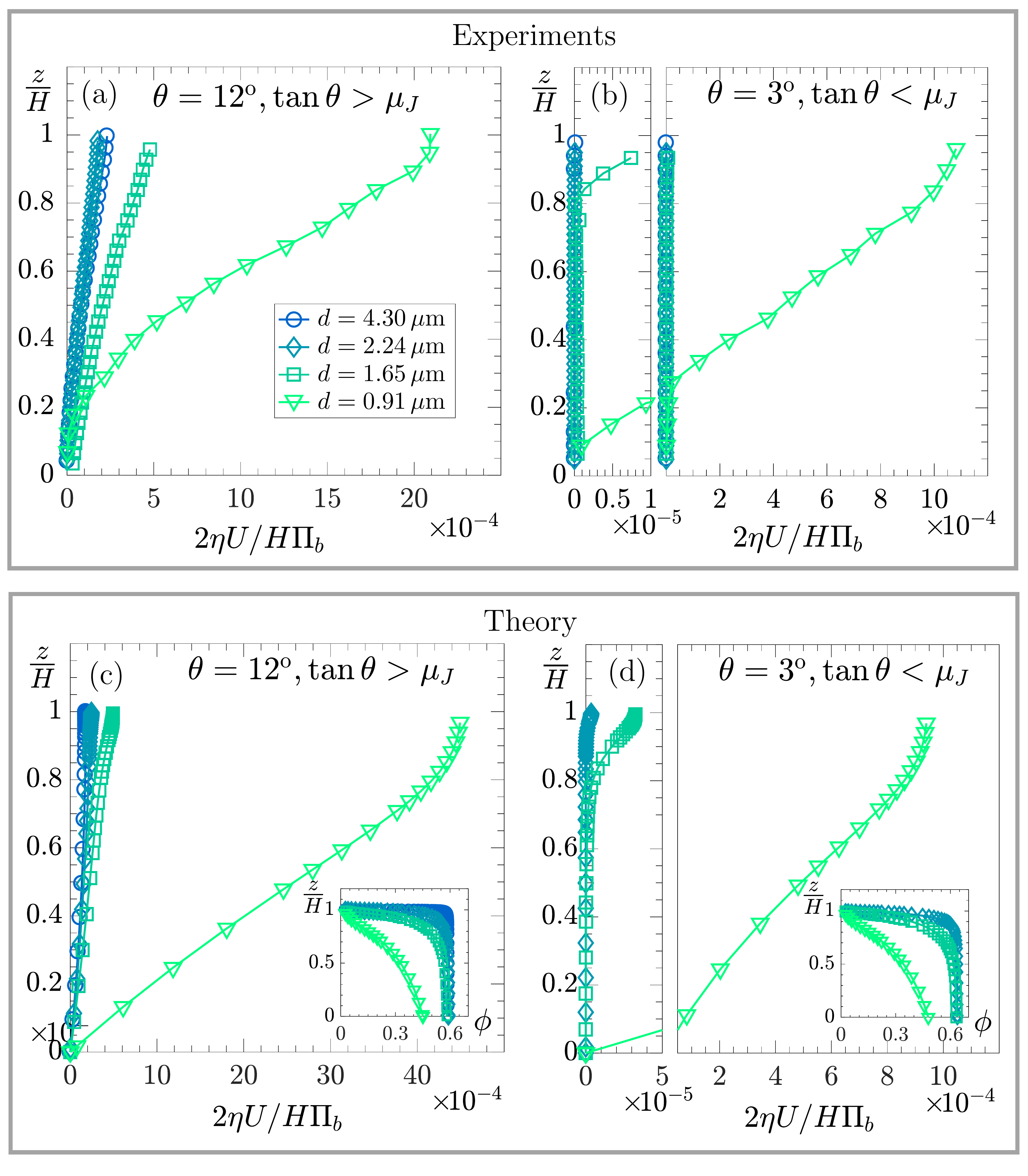}
\caption{{\bf Velocity profiles:}  (a) experimental measurement for $\theta = 12\degree$  and (b) for $\theta = 3\degree$ for the four suspensions; (c) and (d):  corresponding predictions from the theoretical model. insets show the corresponding volume fraction profiles. (on panel (b) and (d) the x-axis is split to zoom in the low velocity range .}
\label{fig:profil}
\vspace{-5mm}
\end{center}
\end{figure}

A closer examination of the flow profiles obtained with $\theta$ respectively above and below the granular athermal angle of repose $\theta_J\simeq 5^{\circ}$ confirms the above observations (Fig.\ref{fig:profil}a and b). In these figures, the dimensionless velocity profile $2 \eta_s U(z)/H\Pi_b$ is plotted as a function of $z/H$, a normalisation that, for athermal systems, gives a velocity profile independent of particle size. In contrats with this athermal prediction, we observe that for $\theta>\theta_J$, the rescale profiles depend on the particles size, with faster flow for smaller particles  (Fig.~\ref{fig:profil}a). This influence of the size is even more dramatic below the angle of repose $\theta<\theta_J$ (Fig.~\ref{fig:profil}b), where large particles do not flow at all, whereas small particles clearly flows over the whole thickness of the layer . For  intermediate particles size, one observes a flow localized at the top of the layer, where the pressure is weaker. 

From the above experimental observation, we conclude that thermal effects promote flows by substantially decreasing the friction coefficient compare to athermal suspensions. For small enough particles, the critical friction coefficient in the quasi static regime vanishes, meaning that particles flows at any inclinations. For intermediate sizes, the friction coefficient is below the athermal friction coefficient but remains finite for the lowest viscous number reached in the experiments. In the following, we develop a theoretical model to describe this phenomenology, based on an additivity model that takes into account the crossover between the glass and  jamming transition.    

\textbf{Model Rheology} -- Our starting point is the additive model introduced in~\cite{ikeda_unified_2012} for the shear stress, to which we add an additive model for the granular pressure in order to get complete rheological description able to describe flow down inclined planes. The shear stress and the granular pressure are written as follows: 
\begin{eqnarray}
\label{eq:addmodel_tau}
\iS(Pe,\phi) =& \iS_s(Pe,\phi) &+ \iS_{\rm th}(Pe,\phi) + \iS_{\rm ath}(Pe,\phi),\\
\label{eq:addmodel_P}
\iP(Pe,\phi) =&  &\;\;\iP_{\rm th}(Pe,\phi)+\iP_{\rm ath}(Pe,\phi),
\end{eqnarray}
where the tilda denotes stresses made dimensionless by the thermal pressure $\sigma_T=k_B T/d^3$. The shear stress (eq.\ref{eq:addmodel_tau}) is composed of three terms ~\cite{ikeda_unified_2012}. The first term  $\iS_s(Pe,\phi) = \eta_s \dot\gamma/\sigma_T = Pe/3\pi$ is the stress stemming from the background solvent, where the P\'eclet number is defined as $Pe = \dot\gamma\tau_T$, with $\tau_T = 3\pi\eta_s d^3 / k_BT$ the thermal microscopic timescale obtained from the Stokes' law. The second term in eq.~\ref{eq:addmodel_tau} is the thermal contribution of the particles to the shear stress, which under the assumption of infinitely hard particles can be expressed as the sum of a yield stress and a shear dependent stress:
\begin{equation}
\label{eq:sheartherm}
\iS_{\rm th}(Pe,\phi)\!=\!\sigma_{GY}(\phi)\!+\!\frac{Y_G}{\left(Pe G(\phi)\right)^{-1}\!+\!\left(1\!+\!p_G Pe^{\alpha_G}\right)^{-1}}
\end{equation}
Below the glass transition, for $\phi<\phi_G$, there is no yield stress and $\sigma_{GY}(\phi)=0$. The function $G(\phi)=h_G(\phi_G-\phi)^{-\gamma_G}$ controls the rapid growth of the relaxation time on approaching the glass transition. For small shear rate, i.e. $Pe\ll G(\phi)^{-1}$, the suspension is a simple Newtonian fluid with $\iS_{\rm th} = Y_G G(\phi) Pe$. At larger $Pe$, $\iS_{\rm th} = Y_G \left(1\!+\!p_G Pe^{\alpha_G}\right)$, with $\alpha_G<1$, describing the onset of the shear thinning plateau related to the slow glassy dynamics. Above the glass transition, for $\phi>\phi_G$, the relaxation time, hence $G(\phi)$, is considered to be infinite, and $\iS_{\rm th} = Y_G + \sigma_{GY}(\phi)+ Y_G p_G Pe^{\alpha_G}$, revealing a yield stress $\iS_{Y} =  Y_G + \sigma_{GY}(\phi)$, with  $\sigma_{GY}(\phi\geq \phi_G)= Y'_G(\phi - \phi_G)^{\beta_G}$(see \cite{ikeda_unified_2012,ikeda_disentangling_2013} for a more detailed discussion). The last term in eq. \ref{eq:addmodel_tau} corresponds to the athermal contribution describing the divergence when approaching the jamming transition, and which is written in the original work of Ikeda et al~\cite{ikeda_unified_2012} as $\iS_{\rm ath}(Pe,\phi)=Y_J K(\phi) Pe$ with \textit{ad-hoc} expression for $K(\phi)$. Here, we rather use the known pressure-imposed rheology of athermal suspension~\cite{guazzelli_rheology_2018} to relate $K(\phi)$ to the friction coefficient $\mu = \iS/\iP$ and the viscous number $J=\eta_s\dot\gamma/\Pi = Pe/3\pi\iP$. We then obtain for the athermal shear stress:

\begin{equation}
\label{eq:shearatherm}
\iS_{\rm ath}(Pe,\phi)= \frac{\mu_{\rm ath}(\phi)}{J_{\rm ath}(\phi)} \frac{Pe}{3\pi},
\end{equation}
with $J_{\rm ath}(\phi)=\left(\frac{\phi_J-\phi}{\phi}\right)^{\gamma_J}$ and $\mu_{\rm ath}(\phi)=\mu_J+bJ_{\rm ath}(\phi)^{1/\gamma_J}$ from the previous work on granular suspensions ~\cite{degiuli2015unified,perrin_thin_2021}.

To complete the rheological description, we must also propose an expression for the granular pressure $\iP$ (eq.~\ref{eq:addmodel_P}). Phenomenological laws have been proposed in the literature for studying particle migration of Brownian suspensions, where the granular pressure was also written as a sum of a thermal and athermal contributions~\cite{frank2003particle,yurkovetsky2008particle}. Here, we use a simplified expression for the thermal pressure component, which assumes a local thermal equilibrium and neglects shear-rate dependence. The thermal pressure is then described by the Carnahan-Starling equation of state for hard spheres~\cite{carnahan1969equation}, modified to take into account the divergence at the jamming transition $\phi_J$~\cite{parisi2010mean}:
\begin{equation}
\iP_{\rm th}(Pe,\phi) = \frac{6}{\pi}\frac{\phi_J}{\phi_J-\phi}\frac{1+\phi+\phi^2-7.5\phi^3}{(1-\phi)^2}\,.
\label{eq:starling}
\end{equation}
For the athermal contribution, we again rely on the known athermal rheology and derive $\iP_{\rm ath}$ using the definition of $J$~\cite{boyer_unifying_2011}: 
\begin{equation}
\iP_{\rm ath}(Pe,\phi) = \frac{1}{J^{\rm ath}(\phi)}\frac{Pe}{3\pi}\,.
\label{eq:athpressure}
\end{equation}

\begin{table}[h!]
\begin{center}
\begin{tabular}{ |c|c|c|c|c|c|c|c|c|c|c|} 
 \hline
 $\phi_G$ 	&  $h_G$ & $\alpha_G$ & $p_G$ & $\beta_G$ & $\gamma_G$ & $Y_G$ & $Y'_G$ & $\phi_J$ & $\gamma_J$ & $\mu_J$ \\ 
 \;0.575 \; &  \; 0.03  \;& \; 0.3  \;& \; 7  \;& \; 0.6  \;& \;  2.2  \;& \; 0.38  \;& \; 0.17  \;& \; 0.64  \;& \; 2.85  \;& \; 0.0875 \\
 \hline
\end{tabular}
\end{center}
\vspace{-3mm}
\caption{Parameters of the model.}
\label{table2}
\end{table}

Equations \ref{eq:addmodel_tau}  and \ref{eq:addmodel_P} with  expressions (\ref{eq:sheartherm}-\ref{eq:athpressure}) provide a phenomenological rheology of a suspension in the thermal crossover, but are expressed considering the volume fraction $\phi$ and P\'eclet number $Pe$ as control parameters. In order to discuss the onset of thermal effects in avalanche flows, it is useful to rewrite the rheology under pressure imposed condition, choosing $\iP$ and $J$ as the control parameters. To do so, one first inverts $\iP(Pe,\phi)$ into $\phi(Pe,\iP)$ and uses $Pe=3\pi J \iP$ to obtain $\phi(J,\iP)$. Then, using these last two expressions in $\iS(Pe,\phi)$ and $\iP(Pe,\phi)$, one finds $\mu(J,\iP)$.

 \begin{figure}[t!]
\begin{center}
\includegraphics[width=14cm]{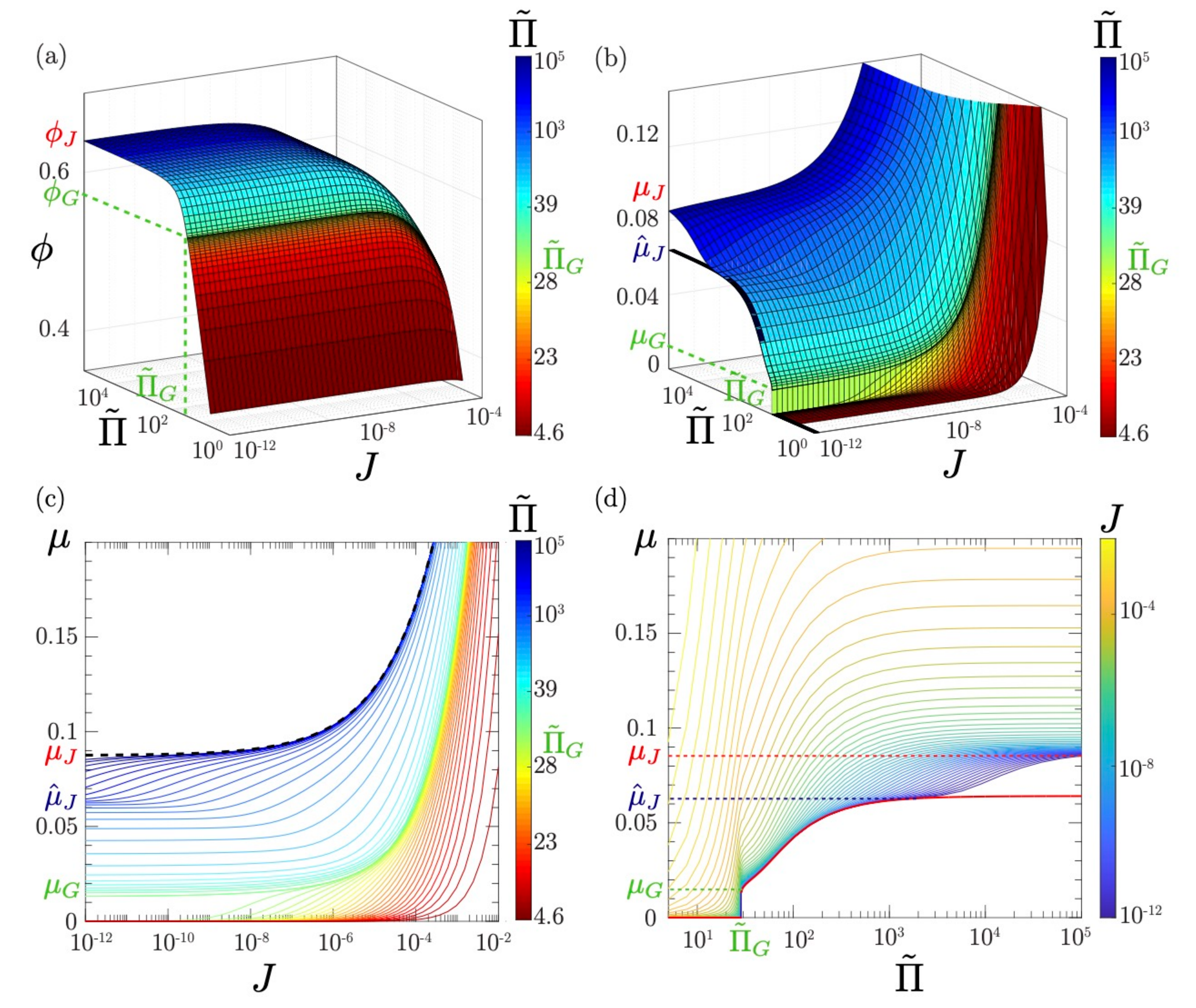}
\caption{{\bf Pressure imposed rheology from the theoretical model:} (a) $\phi(J,\iP)$ and (b) $\mu(J,\iP)$ color coded by $\iP$. (c) $\mu$ as a function of $J$ for different values of $\iP$; (d) $\mu$ as a function of $\iP$ for different $J$. The red curve is $\mu_{J=0}(\iP)$ (see text).}
\label{fig:phimuofJP}
\vspace{-5mm}
\end{center}
\end{figure}

The two constitutive laws $\mu(J,\iP)$ and $\phi(J,\iP)$ hence obtained are plotted as 3D surfaces in Figs.~\ref{fig:phimuofJP}a and \ref{fig:phimuofJP}b, using the model parameters given in table \ref{table2}. The value of  $\phi_G$, $h_G$, $\alpha_G$, $p_G$, $\beta_G$, $\gamma_G$ related to the thermal stress are fixed according to Ikeda et al \cite{ikeda_unified_2012,ikeda_disentangling_2013}. The values of $\phi_J$, $\gamma_J$ and $\mu_J$ are chosen to match the pressure imposed rheology of athermal frictionless particles \cite{degiuli2015unified,perrin_thin_2021}. Parameters $Y_G$ and $Y'_G$ are less constrained in~\cite{ikeda_unified_2012,ikeda_disentangling_2013} and will be fixed to values giving the best fit with the numerical data from Wang and Brady~\cite{wang_constant_2015}, as discussed later in the paper.

Fig. \ref{fig:phimuofJP}c, (resp. Fig. \ref{fig:phimuofJP}d),  shows how the friction coefficient $\mu$ varies as a function of $J$ for different $\iP$, (resp. $\mu$ as a function of $\iP$ for different $J$).  In the limit of large $\iP$, i.e large confining pressure or low temperature, one recovers the athermal behavior and when $J\rightarrow 0$,  $\phi$ converges to $\phi_J=0.64$ and $\mu$ to $\mu_J=0.0875$ (see table \ref{table2}). When introducing thermal effects by decreasing $\iP$, the volume fraction $\phi$ simply decreases and for small values of $\iP$ reaches the limiting value provided by the equilibrium equation of state in absence of flow given by the relation $\iP_{\rm th}(\phi)$ (eq. \ref{eq:starling}). The role of thermal effects on the friction coefficient $\mu$ is more complex. In Fig. \ref{fig:phimuofJP}c, we observe that decreasing $\iP$ decreases the friction coefficient, but that the behavior of $\mu$ when $J$ goes to zero is non trivial, with the emergence of an intermediate plateau below the athermal value $\mu_J$, and a sudden jump to zero when increasing thermal effects. 

%To better understand this transition, we plot in Fig.~\ref{fig:phimuofJP}d in study the behavior of the quasi-static friction coefficient in the thermal crossover  $\mu_{J=0}(\iP)$ plotted as red curve in Fig.~\ref{fig:phimuofJP}d.

To better understand this transition, we focus on the behavior of the quasi-static friction coefficient $\mu_{J=0}(\iP)$ plotted as red curve in Fig.~\ref{fig:phimuofJP}d. This friction coefficient reduces to the ratio of the thermal stresses only, $\tilde{\sigma}_{\rm th}/\tilde{\Pi}_{\rm th}$, because in the limit $J=0$, the P\'eclet number is null (since $Pe=3\pi J\iP$) and athermal stresses vanishes (see eq. \ref{eq:shearatherm}-\ref{eq:athpressure}).  Below the glass transition corresponding to a critical pressure $\iP_G = \iP_{\rm th}(\phi_G)=28.72$, the thermal shear stress for $J=0$ is zero (no yield stress) while the pressure remains finite. Therefore, the quasi-static friction coefficient is null, $\mu_{J=0}(\iP)=0$, and the suspension flows under infinitesimal shear stress.  
Above $\iP_G$, an analytical expression can be derived for $\mu_{J=0}(\iP)$. Approximating the thermal pressure above the glass transition by $\tilde{\Pi}_{\rm th}\simeq 2.6/(\phi_J-\phi)$ (i.e. taking $\phi\simeq 0.6$ for the non-diverging term in the Carnahan-Starling expression) and using $\tilde{\sigma}_{\rm th}(Pe=0,\phi)=Y_G+Y'_G(\phi-\phi_G)^{\beta_G}$ (eq. \ref{eq:sheartherm}), one obtains: 
\begin{equation}
\label{ }
\mu_{J=0}(\iP)=\frac{Y_G}{\iP} +\frac{Y'_G}{2.6^{(1-\beta_G)}} \left(\frac{1}{\iP_G}-\frac{1}{\iP}\right)\,.
\end{equation}

This expression shows that for $\iP=\iP_G$, the friction coefficient jumps to a finite value given by $\mu_G= \mu_{J=0}(\iP_G)=Y_G/\iP_G=0.013$, corresponding to a pile angle of 0.74. The jump is simply related to the  appearance of a finite yield-stress above the glass transition while the pressure remains continuous. Interestingly, the equilibrium pressure at the glass transition sets the characteristic of the transition ($\iP_G$, $\mu_G$) between suspensions with and without a finite critical friction coefficient. When further increasing $\iP$, the quasi-static friction coefficient increases and tends to a constant  $\hat{\mu}_J= \mu_{J=0}(\iP\to \infty)=Y'_G/(2.6^{(1-\beta_G)} \iP_G)$. Using the parameters of table \ref{table2} one get  $\hat\mu_J=0.064$, corresponding to an angle $\hat{\theta}_J=\tan^{-1}(\hat\mu_J)=3.7^{\circ}$, lower than the athermal value  $\theta_J=\tan^{-1}(\mu_J)=5^{\circ}$. In other words, our model predicts that $\mu_J = \lim\limits_{J \rightarrow 0} \mu(J,\iP \rightarrow \infty)$ and $\hat\mu_J = \lim\limits_{\iP \rightarrow \infty} \mu(J \rightarrow 0,\iP)$ differ: the athermal limit is singular, which reflects the existence of both a glass and a jamming transition.  At any given pressure above $\iP_G$, the rheology predicts that a slow flow activated by the thermal fluctuations takes place for $\hat\mu_J<\mu<\mu_J$ but not for  $\mu<\hat\mu_J$.  The consequence of this non trivial prediction for avalanche flows is that a slow creep is expected to occur below the athermal friction angle $ \theta_J=\tan^{-1} (\mu_J)$, but that it will stop at a lower finite angle $\hat{\theta}_J$.  

We conclude this discussion of the model by discussing  the role of the glass yield stress parameters $Y_G$ and $Y'_G$. Whereas $Y_G$ controls the value of the friction coefficient at the glass transition $\mu_G$, $Y'_G$ controls $\hat\mu_J$. Depending on the choice of  these parameters, situations where $\hat\mu_J$ or $\mu_G$ is greater than $\mu_J$ can a priori occur. However, this situation gives rise to non monotonic variation of the friction law, with the appearance of thermally driven hysteresis, not observed in simulations~\cite{wang_constant_2015} or in rotating drum experiments~\cite{berut_brownian_2019}. This shows that the choice of the yield stress parameters considerably modify the shape of the friction law $\mu(J,\iP)$, while they do not qualitatively affect the shape of $\sigma(Pe,\phi)$ curves~\cite{ikeda_disentangling_2013}. The pressure imposed configuration thus provides an interesting approach to access these key parameters of the glass transition. 

\begin{figure}[t!]
\begin{center}
\includegraphics[width=15cm]{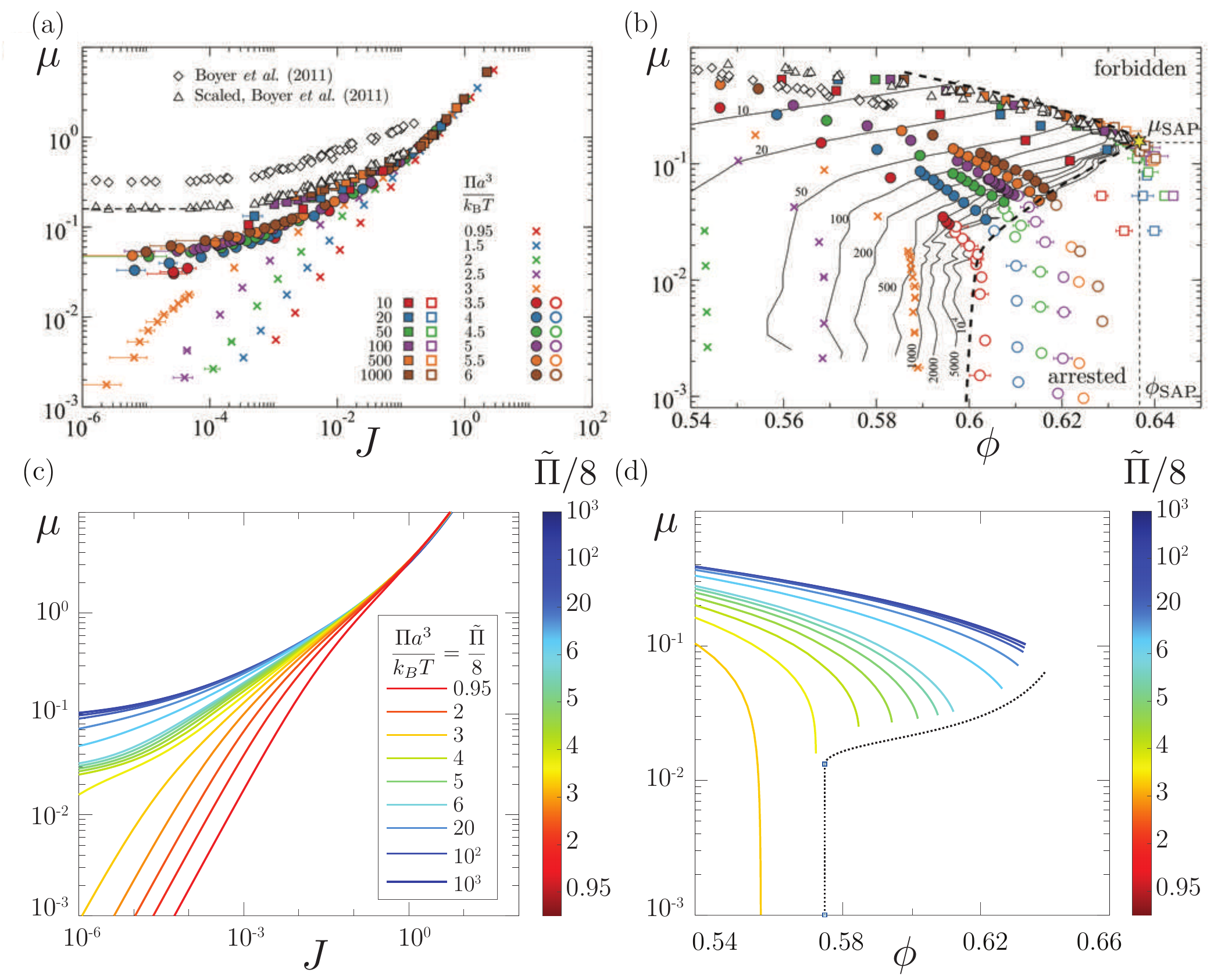}
\caption{{\bf  Comparison with of the model with simulations:} (a) $\mu(J,\iP)$ and (b) $\mu(\phi,\iP)$, Fig. 2 from Wang and Brady paper \cite{wang_constant_2015} (c) and (d) prediction of the model for the same values of $\iP$ as in (a) and (b), the color codes for $\Pi a^3/k_BT=\iP/8$, where $a=d/2$ is the particle radius.}
\label{fig:modelval} 
\end{center}
\vspace{-5mm}
\end{figure}

\textbf{Comparison with numerical simulations and experiments} -- We now validate the model against numerical and experimental data. First, we compare in Fig.~\ref{fig:modelval} our data with the discrete element Brownian simulations done by Wang  and Brady~(figure 2 in \cite{wang_constant_2015}). Fig.~\ref{fig:modelval}a and \ref{fig:modelval}c (resp. Fig.~\ref{fig:modelval}b and d), shows $\mu(J)$ (resp. $\mu(\phi)$) for different $\iP$ obtained  in simulation and predicted by our model. For the set of parameters of table \ref{table2}, the agreement is excellent. 

We next  used the phenomenological additive rheology to compute the steady uniform flow down an inclined plane and compare with our experiments. To do so, we integrate the momentum balance equations $\partial \tilde \sigma / \partial \tilde z+\iP_0\phi \sin \theta=0$ and $\partial \iP / \partial \tilde z+\iP_0\phi \cos \theta=0$, where $\tilde{z}=z/d$ and $\iP_0=\delta \rho g d^4/k_B T$, using  the constitutive law eqs. (\ref{eq:addmodel_tau},\ref{eq:addmodel_P}). This requires a non straightforward inversion of the constitutive law, which is detailed in the Supplementary Material. The predicted velocity profiles using this procedure are plotted in Fig.~\ref{fig:profil}c and \ref{fig:profil}d. Qualitatively, the predicted behavior is similar to the one observed experimentally. However, whereas the case below the athermal repose angle $\theta=3^{\circ}<\theta_J$ (Fig.~\ref{fig:profil}b and \ref{fig:profil}d) is quantitatively captured by the model, for $\theta=12^{\circ}>\theta_J$ (Fig. \ref{fig:profil}b and d) a factor two exists between the predicted velocity and the measured one. This discrepancy may come from side wall effects or non local effect present in the experiments and not taken into account in the model. The model also gives access to the packing fraction profile (inset of Fig.~\ref{fig:profil}b and \ref{fig:profil}d), showing that thermal effects induce significant density gradient close to the interface, an effect similar to the Perrin equilibrium density distribution~\cite{perrin1913atoms}. Similar qualitative trends are observed in the experiments but we have not been able to perform quantitative volume fraction measurements, due to light inhomogeneity.  From the computed velocity profiles, we can compute the effective viscous number $J_{\rm eff}$ as in the experiments,  and compared the predicted (light color disks) and experimental (plain color symbols) effective rheology $\mu(J_{\rm eff},\iP_b)$ (Fig.~\ref{fig:effrheo}). The agreement is remarkably good considering the assumptions made in the additive model.   

\textbf{Conclusion} -- In this paper, we have revisited the classical configuration of granular flows on an inclined plane for microscopic particles, in a regime where thermal effects start to play a role. Using a confocal inclined microscope combined with miniaturized set-up, we were able to experimentally extract an effective pressure imposed  rheology from the velocity profile measurements. By varying the particles size, we were able experimentally to control the dimensionless parameter that controls thermal agitation in this configuration, i.e. the  ratio of the particle pressure due to gravity to the thermal pressure: $\iP=\Pi d^3/k_BT$.    At high agitation (low $\iP$), we found that the material flows for infinitely small slopes, meaning that the macroscopic friction coefficient vanishes, whereas at small agitation (high $\iP$), we recover the athermal behavior with a finite repose angle given by the friction coefficient at the jamming transition $\mu_J\simeq 0.1$ for athermal frictionless spheres. Inspired by Ikeda et al~\cite{ikeda_unified_2012}, we developed a phenomenological model based on the sum of a thermal contribution describing the glass transition and an athermal contribution capturing the jamming transition, which provides the pressure imposed constitutive law for thermally agitated granular media  $\mu(J,\iP)$ and $\phi(J,\iP)$. The model reproduces well the experimental observations. A major prediction of the model is that the quasi-static friction coefficient when decreasing the thermal activity (increasing $\iP$) suddenly jumps from zero to a finite value when crossing the glass transition, and tends to a value below the athermal angle of friction. Therefore, the signature of the glass transition in the framework of a pressure imposed rheology is the appearance of a glassy friction angle whose value is distinct from the jamming friction angle. Characterizing experimentally this hypothetical glassy friction angle is a challenge. A possibility would be to study the relaxation of weakly Brownian heaps over very long periods of time and check whether the pile eventually stops at a finite angle below the athermal angle. Beyond the question of the threshold, our model  provides the rheology of Brownian granular media in the full range of viscous number and can thus be used to predict flow in other configurations such as avalanches in rotating drums or flow in silos.

%
%\begin{eqnarray}
%\frac{d\phi}{dz} &=& -\iP_0 \cos\theta \chi\left[\phi(z)\right] \phi(z)
%\quad {\rm with} \quad \chi =\left(\frac{d\iP}{d\phi}\right)^{-1}\\
%\iP(\phi)&=& \iP(\phi,J_{\mu_0}(phi))
%\quad {\rm with} \quad \mu(\phi,J_{\mu_0}(\phi)) = \mu_0  		
%\label{eq:diff}
%\end{eqnarray}
%

%\begin{figure}[b!]
%\begin{center}
%\includegraphics[width=\columnwidth]{Figures/Fig7.pdf}
%\caption{{\bf Effective friction in the thermal crossover:} (a) $\mu(J)$ for a set of $\iP$, color coded by $\iP$ and (b) $\mu(\iP)$, for a set of $J$ color coded by $J$.}
%\label{fig:muofJP}
%\vspace{-5mm}
%\end{center}
%\end{figure}
%
%
%

%From this we extract an effective rheology and link it to a previous  theoretical model which we enrich for the sake of our study. Going further in the analysis, we could look in more details at the shapes of the velocity profiles. However, wall effects, non-locality or frictionless-frictional transition as well as the difficulty to measure small velocities hinder a fruitful comparison with theoretical predictions. [{\it Except for the small grains which are less sensible to the wall effects? The shape of the velocity profiles of the small grains corresponds quite well to the predictions.}]. Future works could improve the experimental setup in order to access the flow further away from the wall, by tuning more finely the optical index of the suspension. It would also be very interesting to confront the pressure-imposed model to other typical pressure-imposed configurations rotating drum experiments.  the orle of friction. 

%\bibliographystyle{apsrev4-1}
\bibliography{bgs.bib}

\end{document}